\documentclass[%
 reprint,
 jmp,
superscriptaddress,
 amsmath,amssymb,
 aps,
]{revtex4-2}

\usepackage{graphicx}% Include figure files
\usepackage{dcolumn}% Align table columns on decimal point
\usepackage{bm}% bold math
\usepackage{svg}
\usepackage{xcolor}
\begin{document}

\preprint{APS/123-QED}

\title{Selective linewidth control in a micro-resonator\\ with a resonant interferometric coupler}
\author{Paula L. Pagano}
\affiliation{
Centro de Investigaciones Ópticas, CONICET-CICBA-UNLP, Camino Centenario y 505 s/n , 1897 Gonnet, Argentina.}
\affiliation{Dipartimento di Ingegneria Industriale e dell'Informazione, Università di Pavia, Via Ferrata 5, 27100 Pavia, Italy.}
\author{Massimo Borghi}
\email{corresponding author: massimo.borghi@unipv.it} 
\author{Federica Moroni}
\author{Alice Viola}
\author{Francesco Malaspina}
\affiliation{Dipartimento di Fisica, Università di Pavia, Via Bassi 6, 27100 Pavia, Italy.}

\newcommand\AM[1]{\textcolor{black}{#1}}
%\author{Matteo Menotti}
%\affiliation{Xanadu Quantum Technologies, 2400-777 Bay St., Toronto, ON M5G 2C8, Canada.}

\author{Marco Liscidini}
\affiliation{Dipartimento di Fisica, Università di Pavia, Via Bassi 6, 27100 Pavia, Italy.}

\author{Daniele Bajoni}
\affiliation{Dipartimento di Ingegneria Industriale e dell'Informazione, Università di Pavia, Via Ferrata 5, 27100 Pavia, Italy.}

\author{Matteo Galli}
\affiliation{Dipartimento di Fisica, Università di Pavia, Via Bassi 6, 27100 Pavia, Italy.}

%

%  \homepage{http://www.Second.institution.edu/~Charlie.Author}


\date{\today}% It is always \today, today,

\begin{abstract}
Optical microresonators are characterized by a comb of resonances that preserve similar characteristics over a broad spectral interval. However, for many applications it is beneficial to selectively control of the quality factor (Q) of one or only some resonances. \\
In this work we propose and experimentally validate the use of a resonant interferometric coupler to selectively change the Q-factor of a target resonance in an integrated silicon nitride microresonator. We show that its Q-factor can be continuously tuned from $6.5\times10^4$ to $3\times10^6$, leaving the untargeted resonances uperturbed. Our design can be scaled to independently control several resonances.

\end{abstract}

\maketitle

%\tableofcontents

\section{\label{sec:intro} Introduction}
Integrated optical microresonators are versatile devices which find use in several applications, such as spectral filters, switches \cite{qiang2007optical,huang2020optical,mokhtari2021tunable, selim2023enhanced}, delay lines \cite{lin2022high,shan2021broadband}, sensors \cite{steglich2021silicon, kazanskiy2023review}, modulators \cite{moradi2021design, hagan2020high} and light sources \cite{foster2011silicon}. Microresonators are typically characterized by a comb of resonances that preserve similar characteristics, such as the quality factor or the extinction rate, over a broad spectral interval. This feature is harnessed to generate broadband frequency combs \cite{shen2020integrated}, soliton pulses \cite{kippenberg2018dissipative}, or photon pairs in high dimensional entangled states \cite{kues2017chip}. However, some applications may require the selective and dynamic control of the quality factor ($Q$) of only some of the resonances. Examples include $Q$-switched lasers \cite{lei2019numerical, shtyrkova2019integrated}, single frequency lasers \cite{miller2015tunable}, tunable modulators \cite{shoman2019compact}, tunable delay lines \cite{liu2018ultra}, optical memories \cite{zhang2021proposal}, pseudo-random binary sequence generators \cite{rakshit2021proposal} and photon pair sources with tailored spectral correlations \cite{vernon2017truly} or high heralding efficiency \cite{chen2023pushing,burridge2023integrate}. Several approaches have been investigated to control the Q-factor \cite{wen2012all,wen2011all,guo2021all,shoman2019compact, chen2007compact,zhang2021proposal, miller2015tunable, manipatruni2008high, burridge2023integrate,zhang2021proposal, miller2015tunable, manipatruni2008high, xue2015normal}, and some of them are wavelength selective \cite{wen2012all,wen2011all,burridge2023integrate,zhang2021proposal, miller2015tunable, manipatruni2008high, xue2015normal}. For example, a strong control pulse has been used to induce Raman scattering in silicon on a target resonance \cite{wen2012all}. However, this operation can be done efficiently only at the maximum of the Raman gain of the material, and demands strict fabrication tolerances \cite{wen2011all,wen2012all}. %The coupling coefficient of the directional coupler of a ring can be changed using resistance tracks, exhibiting a Q factor tunable between 9,000 and 96,000 \cite{strain2015tunable}. Nevertheless, the target resonance shifts as the Q factor is detuned. 
The Q-factor can also be tuned by varying the phase shift in an unbalanced interferometer forming the coupling region of a microresonator \cite{shoman2019compact, chen2007compact}. Alternatively, the Q-factor of a single resonance can be changed due to the strong coupling with another resonator \cite{zhang2021proposal, miller2015tunable, manipatruni2008high, xue2015normal}. 
or by a combination of the previous strategies \cite{burridge2023integrate}.\\
%In the second case, the Q factor can be tuned using thermo-optical effect. For example, a device with graphene-on-silicon, where a graphene sheet covers the nanobeam cavity with a U-shaped pattern \cite{guo2021all}. 
In this work, we propose and experimentally demonstrate the selective control of the Q-factor of a single resonance in a microresonator by using a  resonant interferometric coupler. A Main resonator is coupled twice to a bus waveguide to form an unbalanced Mach-Zehnder interferometer (MZI). Then, we introduce an auxiliary resonator (Aux) in one of the arms to impart a wavelength-selective phase. This changes the interference within the MZI and thus the effective coupling coefficient with the input waveguide. The article is structured as follows: in Sec.\ref{sec:theory} we describe the working principle of the device and explore different configurations which allows us to selectively change the Q-factor of a target resonance. The device is then fabricated on a silicon nitride chip, and its experimental characterization is reported in Sec.\ref{sec:device}. Here we show that the Q-factor of the target resonance can be continuously changed from $6.5\times10^4$ to $3\times10^6$ with minimal perturbation of the adjacent ones. 

%This work is organized as follows: after describing in Sect.\ref{sec:intro} the state-of-the-art and the approaches followed to achieve control of the Q factor, in Sect.\ref{sec:theory} a description of the device is developed exploring possible working regimes, Sect.\ref{sec:device} describes how the device was fabricated and the experimental set-up used for its characterization, Sect.\ref{sec:linear_beh} presents the experimental characterization of the device,
%with balanced coupler and over-coupled auxiliary resonator
%the concluding remarks and possible applications of the device are delivered in Sect.\ref{sec:concl}

%When the couplers are asymmetric, destructive interference never leads to under-coupler regime. Having the auxiliary resonator over-coupled add a phase of $\pi$ to the system. Meanwhile if it is critically coupled, when it is on resonance, it has 0 transmission those breaks the interferometer. Moreover, the response of those systems is analyzed while the auxiliary resonator is detuned. Finally, the experimental characterization of a device with balanced coupler and over-coupled auxiliary resonator is presented on a silicon nitride photonic chip. 

\section{\label{sec:theory}  Working principle of the device}
\begin{figure}[t!]
    \centering
    \includegraphics[width = \columnwidth]{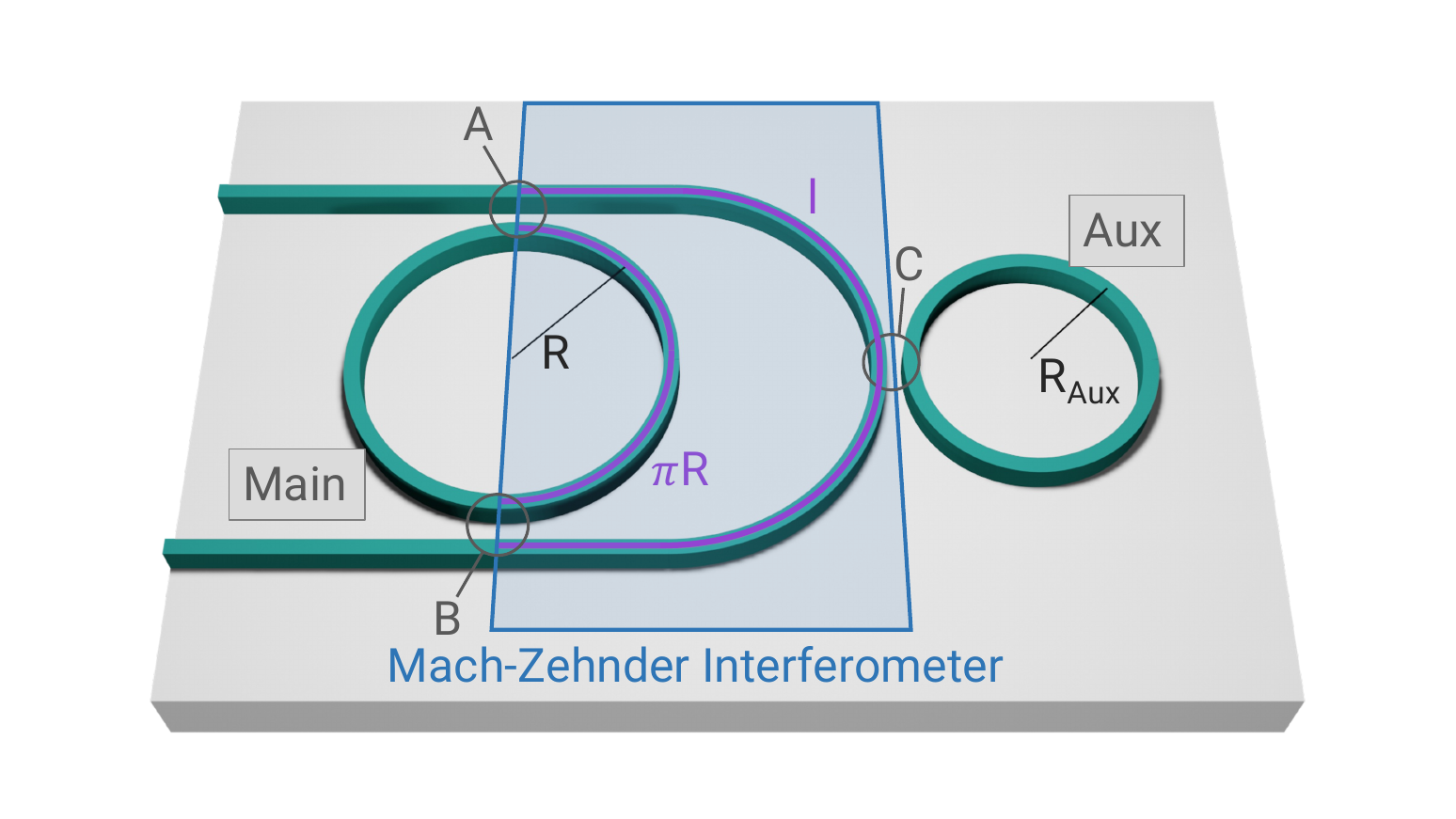}
    \caption{Sketch of the proposed device. Waveguides are shown in green, while the lower cladding region is shown in gray.}
    \label{fig:sketch}
\end{figure}

%version Paula:
The proposed structure is sketched in Fig.\ref{fig:sketch}. A main ring resonator (Main) of radius $R$ is coupled twice to a bus-waveguide \AM{at points A and B, with coupling coefficients $\kappa_A$ and $\kappa_B$ respectively,} to form an interferometric coupler \cite{shoman2019compact}.
An auxiliary resonator of radius $R_{\rm{Aux}}$ is also coupled to the bus waveguide %PAULA'S VERSION: one of the interferometer arms
at point C. 
% Alice's version
% The device is sketched in Fig. \ref{fig:sketch}. It is composed of a Main ring resonator that is coupled twice (at points A and B respectively) to a bus-waveguide to form an interferometric coupler. 
The highlighted area in Fig.\ref{fig:sketch} can be seen as a Mach-Zehnder interferometer (MZI), with a first arm of length $\pi R$ along the Main ring, and a second arm of length $l$ along the bus-waveguide.

The effective coupling of the Main ring resonator to the bus waveguide depends on the phase difference between the two MZI arms \cite{shoman2019compact, chen2007compact, gentry2016CLEO}. 
Such a difference is given by two distinct contributions: the imbalance in the optical length of the two interferometer arms, and the transmission of the Aux ring.
The former contribution acts at all frequencies, the latter is relevant only when its frequency is within the spectral width of an Aux ring resonance.
Thus, one has two sets of resonances that can be controlled independently: those that belong only to the Main ring and those that are shared with the Aux.
For instance, by choosing incommensurate values for $R$ and $R_{Aux}$, it is possible to selectively control one single resonance. Using a microheater, one can adjust the refractive index of the bus waveguide to set the desired coupling for all the other Main resonances.  An additional microheater can act on the Aux to align one of its resonances to the target Main ring resonance.

With reference to Fig.\ref{fig:sketch}, the device transmission $T$ at the output port can be written as 
\begin{equation}
    T = \left|\rm{e}^{i\tilde\phi_{\rm{Main}}}  \frac{\sigma_A\sigma_B  \delta_{\rm{MZI}} - \kappa_A\kappa_B - \delta_{\rm{MZI}}e^{i2\tilde\phi_{\rm{Main}}}}{1 - \left[\sigma_A\sigma_B - \kappa_A\kappa_B  \delta_{\rm{MZI}}\right] e^{i2\tilde\phi_{\rm{Main}}}}\right|^2,
    \label{eq:T}
\end{equation}
where $\sigma_j$ and $\kappa_j$ are the self- and cross-coupling coefficients at point $j \in \{A, B, C\}$, $\tilde\phi_{\rm{Main}} = \tilde\beta \pi R$  and $\tilde\beta = \beta + i\frac{\alpha}{2}$ is the complex wavevector, which accounts for both propagation ($\beta$) and linear attenuation ($\alpha$). Finally,
\begin{equation}
    \delta_{\rm{MZI}} = t_{\rm{Aux}} \rm{e}^{i\tilde\beta \left(l -\pi R\right)}
\end{equation}
describes the attenuation %amplitude
and phase difference acquired between the two arms of the MZI, and
\begin{equation}
    t_{\rm{Aux}} = \frac{\sigma_C - \rm{e}^{i\tilde\phi_{\rm{Aux}}}}{1 - \sigma_C e^{i\tilde\phi_{\rm{Aux}}}},
\end{equation}
is the transmission amplitude of the Aux resonator, with $\tilde\phi_{\rm{Aux}} = \tilde\beta 2\pi R_{\rm{Aux}}$.

%. Light propagating in the waveguide or in the ring through a section of length $L$ is subject to a dephasing $\beta L$ and attenuated by a factor $e^{-\alpha L}$, where $\beta$ is the wavevector and $\alpha$ is the attenuation. We account for all of this by considering a complex dephasing
%\begin{equation}
%    \tilde\phi = (\beta+i\alpha)L =\tilde{\beta}L,
%\end{equation}
%where $\tilde\beta = \beta+i\alpha$ is a complex wavevector. For instance, light propagating in the Main ring between points A and B is dephased by $\tilde\phi_R = \tilde\beta \pi R$. Another important quantity is the phase difference $\Delta\phi_{\rm{MZI}}$ between the two arms of the MZI.
%Points A, B, and C are described as point couplers characterized by amplitude cross-coupling coefficients $\kappa_A$, $\kappa_B$, and $\kappa_C$ respectively. We find that 
Unsurprisingly, Eq.(\ref{eq:T}) is equivalent to the transmission of a resonator in the all-pass filter configuration in which the effective coupling coefficient $\kappa_{\textup{eff}}^2$ with the input waveguide is given by 
\begin{equation}
|\kappa_{\textup{eff}}|^2= |\kappa_A\sigma_B +\kappa_B\sigma_A\delta_{\textup{MZI}}|^2, \label{eq:keff}     
\end{equation}
which is the expression for the cross port transmission of the MZI, where $\Delta\phi_{\rm{MZI}} = \beta \left(l -\pi R\right) + \arg[t_{\rm{Aux}}]$. 

We now consider the special case in which $l = 3\pi R$. Thus, for all of the resonances that are spectrally far from those of the Aux (for which $|\delta_{\textup{MZI}}|\sim 1$ and arg$[t_{\textup{Aux}}]\sim 0$) we have that the phase difference in the MZI is a multiple of $2\pi$, and the interference is constructive at the cross port of the MZI. In this situation, when $\kappa_A = \kappa_B$, the effective coupling is maximum and equal to $|\kappa_{\textup{eff}}|^2 = 4\kappa_A^2\sigma_A^2$. Changing $\Delta\phi_{\rm{MZI}}$ through a micro-heater placed in the long arm of the MZI modifies $\kappa_{\textup{eff}}^2$ for all the resonances. However, if the Aux resonator is used to change the interference condition at the MZI, the effective coupling is modified only in the neighborhood of the Aux resonances. It is then possible to vary the effective coupling of the Main ring from a completely uncoupled condition, when interference is destructive, to an over-coupled condition, when the interference is constructive. %Depending on the application, one can set a minimum coupling condition different than undercoupling, by considering $\kappa_A\neq\kappa_B$.
%This is the situation that realizes the largest effective coupling ($2\kappa_A$) of the Main resonator to the bus-waveguide.  We notice that in a practical realization, one can envision a tuning mechanism, such as a micro-heater, to change $\Delta\phi_{\rm{MZI}}$ and tune the effective coupling for all the resonances in the same way. In this case, one can vary the effective coupling to the Main ring from a completely uncoupled condition when interference is destructive to overcoupling when the interference is constructive. Depending on the application, one might also want to set a minimum coupling condition different than undercoupling. This can be obtained by considering $\kappa_A\neq\kappa_B$.

We further set $R_{\rm{Aux}}=\frac{3}{4}R$ so that only one resonance out of four in the Main ring will overlap with one of the Aux, leaving the other resonances largely detuned from any Aux resonance. We consider the case in which the Aux is massively over-coupled, so that most of the light is efficiently coupled into it and back to the MZI before it is lost by scattering. 

%Finally, take $R_{\rm{Aux}}=\frac{3}{4}R$ such that only one every fourth resonance of the Main ring overlaps with one of the Aux, while the others are well detuned from any Aux resonance. We consider the case in which the Aux is heavily over-coupled ($\kappa_C=0.1$) so that most of the light is efficiently coupled into it and back to the MZI before it is lost by scattering. 

\begin{figure*}[hbt!]
    \centering
    \includegraphics[width=\textwidth]{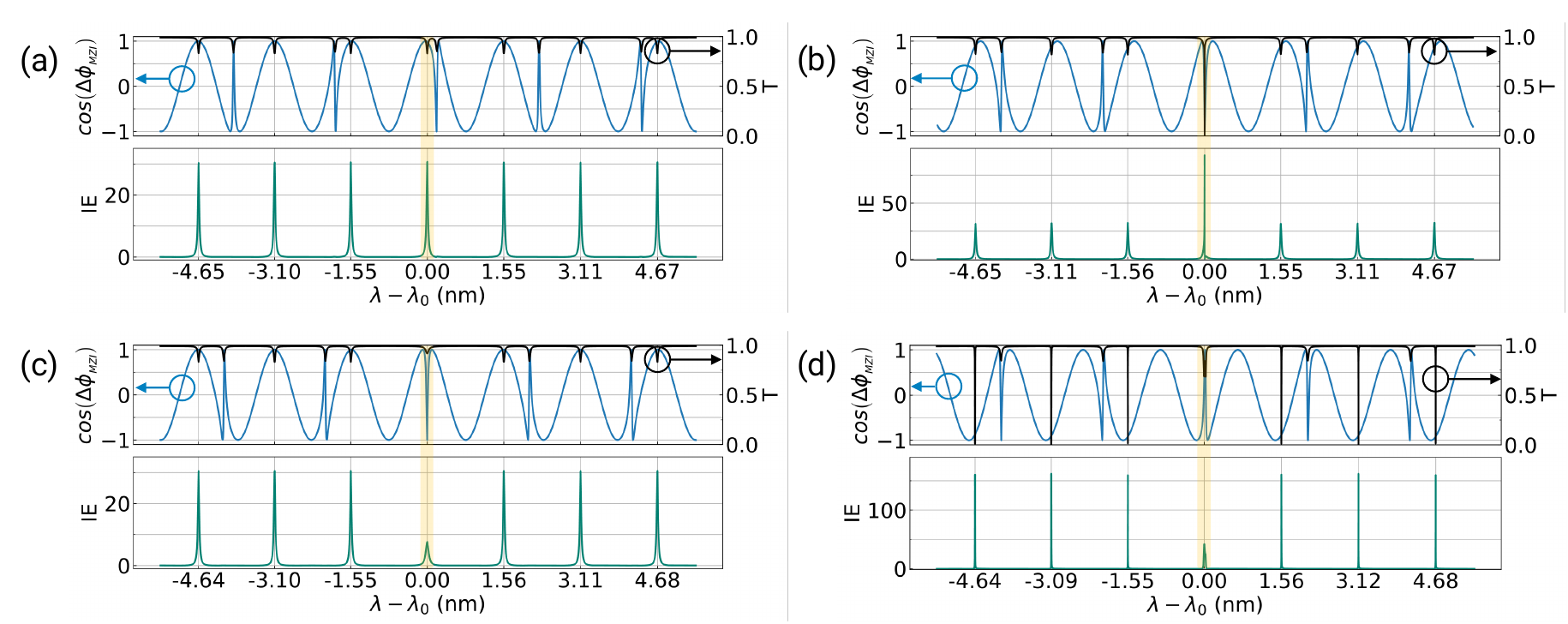}
    \caption{Cosine of the MZI dephasing %phase
    $\Delta\phi_{\textup{MZI}}$ (blue), normalized transmission $T$ (black), and intensity enhancement spectra (green) as a function of the detuning from a target resonance at wavelength $\lambda_0$ (highlighted in yellow). In this simulation, we set $\kappa_A^2 = \kappa_B^2=0.03$, $R=116.6\,\mu m$, $\alpha = 0.1\,$ dB/cm and $\kappa_C = 0.1$. (a) $\Delta\phi_{\textup{MZI}}=2m\pi$ and the Aux resonator is detuned from all the Main resonances. (b) Same as in (a), but the Aux resonator is tuned to match the highlighted resonance. (c) The Aux ring turns the target resonance to critical coupling when one of its resonances is tuned at $\lambda_0$. (d) The MZI fringes are detuned to have a comb of critically coupled Main resonances, while the Aux resonator overcouples the target resonance.} %When the Aux is detuned from all of the resonances (a), they have similar intensity enhancements. When one Aux resonance matches one resonance of the Main (b-c-d), the latter behaves very differently from the other Main resonances. This method can be used for instance to have an uncoupled (b) or critically coupled (c) resonance in a comb of heavily over-coupled ones, or an over-coupled resonance in a critically-coupled comb (d). Note the different intensity enhancement scales in the plots (c-d) compared to plots (a-b).}
    \label{fig:IE}
\end{figure*}
In Fig.\ref{fig:IE} we show the spectral transmission $T$, the cosine of $\Delta\phi_{\rm{MZI}}$, and the intensity enhancement (IE) in the Main ring for different MZI and Aux configurations. The spectral position of the Aux resonances can be clearly seen as sharp variations in the otherwise smooth sinusoidal oscillations of $\cos(\Delta\phi_{\rm{MZI}})$. In Fig.\ref{fig:IE}(a), $\beta (2\pi R) = 2m\pi$, so $\kappa_{\textup{eff}}^2$ is maximized and the Main ring is maximally over-coupled. The Aux is detuned from all the Main resonances, which are then unperturbed and have the same value of the IE.  %On the other hand, they can barely be identified in the transmission and intensity enhancement spectra due to their massive overcoupling.
In Fig.\ref{fig:IE}(b) we show the response of the structure with an Aux resonance matching a target Main resonance at the wavelength $\lambda_0$, which is highlighted in yellow. The Main ring becomes completely uncoupled at $\lambda_0$ due to the additional phase imparted by the Aux resonator, which causes $\Delta\phi_{\rm{MZI}}(\omega_0)=\pi$ and sets the MZI to a completely destructive interference at its cross port. %From the IE we note that, despite the undercoupling condition, the linewidth seems to be broader than the one of the neighbouring resonances. This is because the transmission dip is the result of the overlap between the Main and Aux resonances.% so it can not be treated simply as a Lorentzian lineshape.

Figure \ref{fig:IE}(c) shows the effect of a slight detuning of the MZI from the configuration shown in Fig.\ref{fig:IE}(b). Here $\Delta\phi_{\rm{MZI}}>0$ for all the resonances, and the $\pi$ phase shift given by the Aux does not uncouple the target resonance anymore because it causes $\Delta\phi>\pi$. %Again, the target peak is not exactly Lorentzian. 
We choose the MZI detuning such that the target resonance is brought to a critical coupling condition when it is overlapped by the Aux ($\kappa_{\textup{eff}}=\kappa_{cc}$, where $\kappa_{cc}$ is the coupling coefficient for which the intrinsic loss perfectly balances the coupling loss with the input waveguide), while the other ones are still over-coupled. This condition is also clearly witnessed from the large IE at the target resonance.%although not to the maximal value.

In the last configuration, reported in Fig.\ref{fig:IE}(d), we show that reverse behavior is possible as well: a target resonance in a comb of critically coupled ones can be brought to overcoupling due to the use of the Aux resonator. The IE at the target resonance is now greatly reduced compared to the neighboring ones.
In Sec.\ref{sec:device} we will show that the Aux resonator can be swept continuously across the Main resonance, thus realizing all the coupling conditions described above. %configurations discussed above.

%In section \ref{sec:linear_beh} we will show that the Aux can be swept continuously across the Main resonance, making it possible to reach all the possible coupling conditions.

%\begin{figure}[hbt!]
%    \centering
%    \includegraphics[width=\columnwidth]{WatchOut?.png}
%    \caption{\color{brown}This is to show the original configuration of the resonant interferometric coupler, to show the importance of the phase difference in the MZI and how the non-lorentzianity of the peaks can lead to misinterpretations of the spectra. The target resonance is completely under-coupled, as is clear from the fact that $\Delta\phi_{\rm{MZI}}=\pi$. Still its linewidth seems to be broader than the one of the neighbouring resonances. This is because the peak is the result of the overlap between two non-lorentzian peaks. SHOULD I SHOW THIS?}
%    \label{fig:?}
%\end{figure}
\section{\label{sec:device} Experimental results}
\begin{figure}[b!]
    \includegraphics[width = \columnwidth]{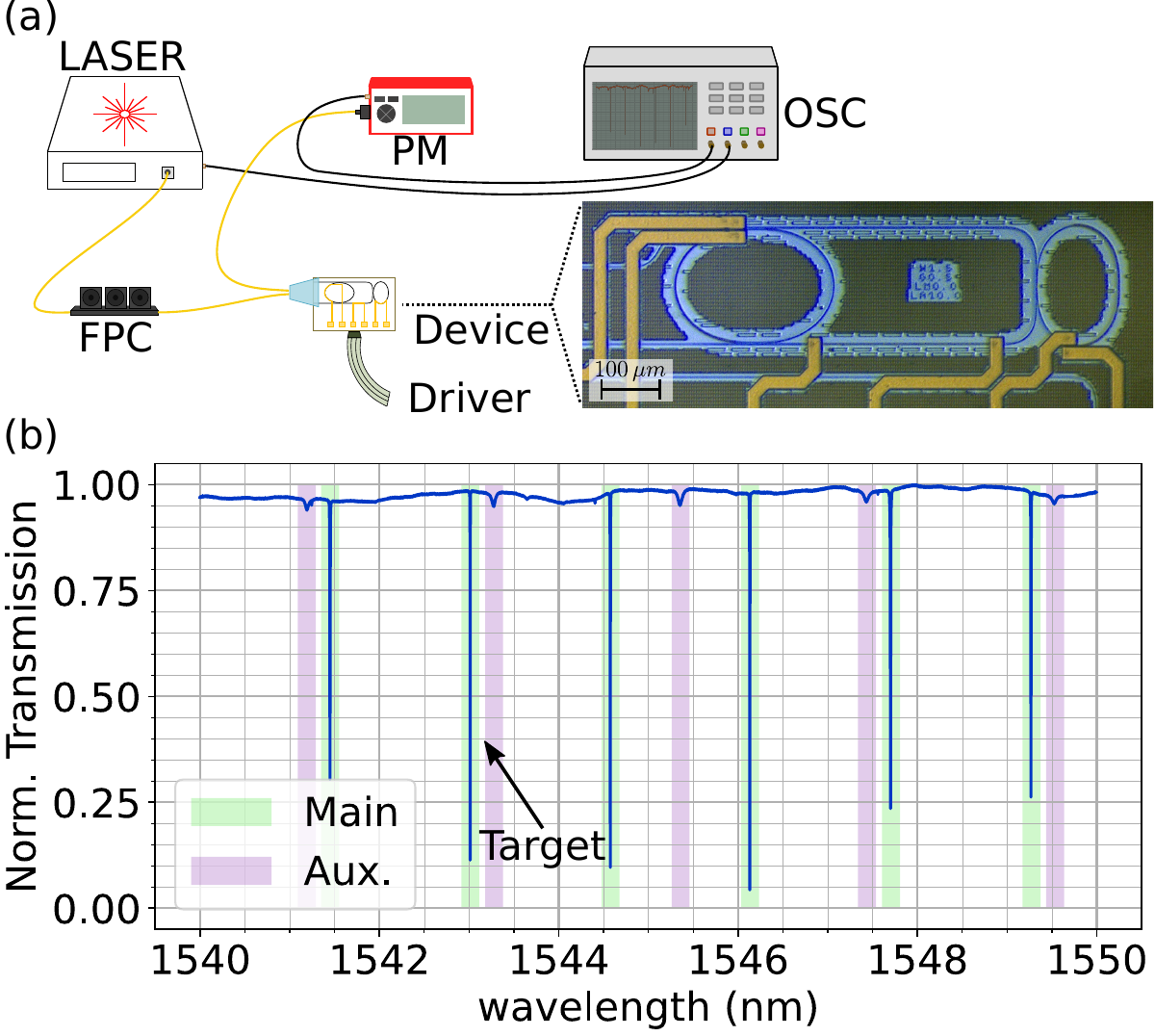}
    \caption{(a) Sketch of the experimental setup FPC: fiber polarization controller, PM: power meter, OSC: oscilloscope. The right-inset shows an optical microscope image of the device. (b) Transmission spectra of the device. A resonance of the Main resonator is highlighted in green, while a resonance of the Aux resonator is highlighted in violet. The arrow indicates the target resonance at $\lambda_0=1542.97$ nm}
    \label{fig:picture}
\end{figure}

The device described in Sec.\ref{sec:theory} is fabricated on a silicon nitride photonic chip. An optical microscope image is shown in Fig.\ref{fig:picture}(a). Three metallic heaters allow us to respectively change the resonance wavelength of the Main ring, of the Aux resonator, and the phase of the bus-waveguide arm $\phi_{\textup{wg}}=e^{i\beta d}$ through the thermo-optic effect.
The sample is mounted on a holder that is thermally stabilized, and the metallic micro-heaters are controlled by a voltage driver module. 
The waveguide has width of \mbox{1.75 $\mu$m} and  thickness of \mbox{800 nm}, while the gap between the Main resonator and the bus-waveguide is \mbox{0.4 $\mu$m}.
The setup used to probe the device transmission $T$ is shown in Fig. \ref{fig:picture}(a). A tunable laser is coupled to a fiber polarization controller (FPC) and the polarization is controlled to match that of the TE mode of the silicon nitride waveguide.
%with its polarization regulated by a fiber polarization controller (FPC). 
Light is coupled into and out of the chip through a fiber array of UHNA4 fibers with a coupling loss of approximately 1.7 dB/facet. Then, the output signal is detected by a powermeter and recorded in real-time by an oscilloscope. 
\begin{figure*}[t!]
    \centering
    %\includegraphics[width = \textwidth]{4-spectra.png}
    %\includesvg[width = \textwidth]{four-stack.svg}
    \includegraphics[width = \textwidth]{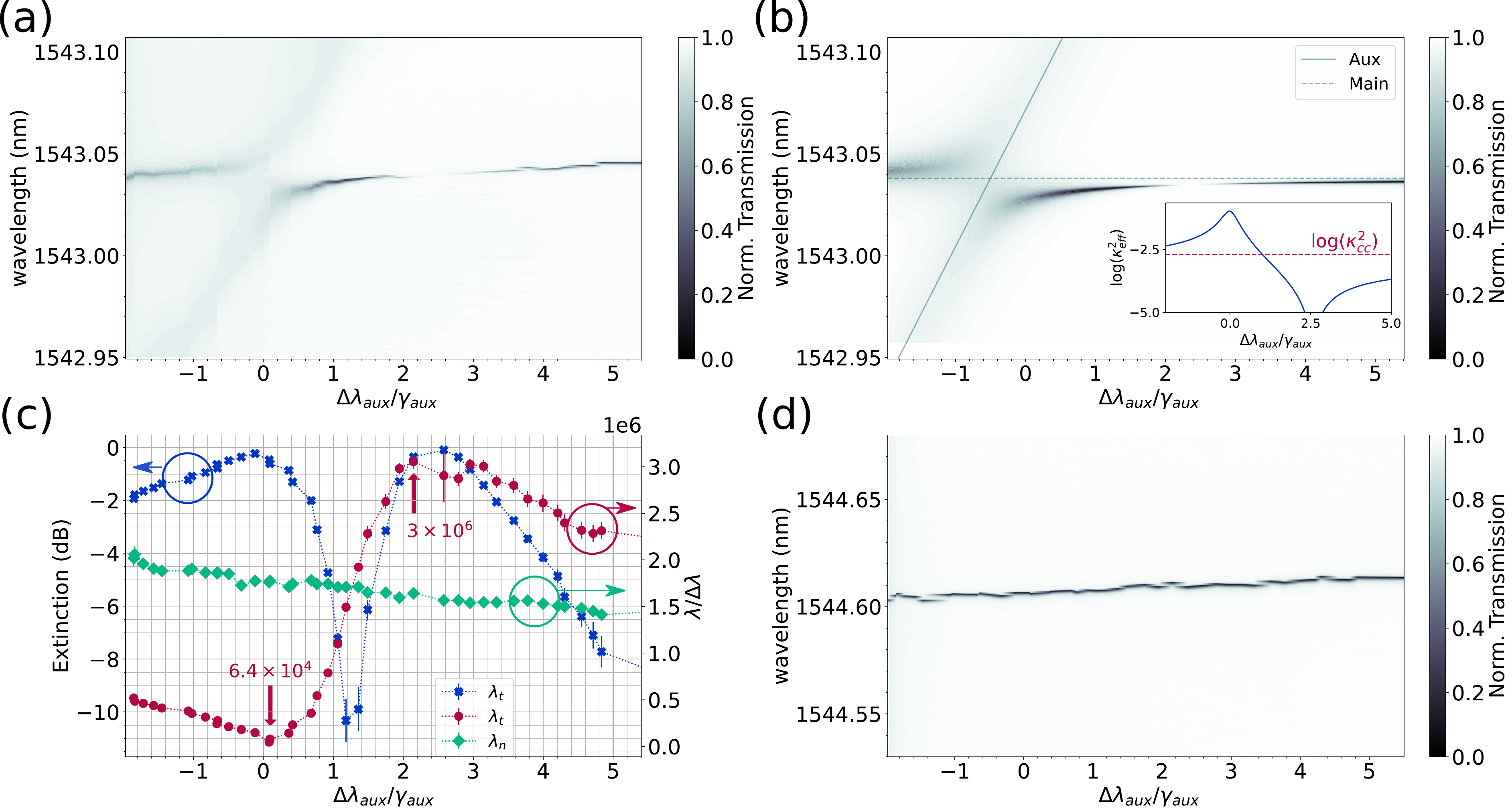}
    \caption{(a) Stacked spectra of the target Main resonator  resonance (see Fig.\ref{fig:picture}(b)) for different normalized wavelength detunings of the Aux resonator $\Delta\lambda_{\textup{aux}}/\gamma_{\textup{aux}}$. (b) Simulation of the device transmission spectra $T$ using Eq.(\ref{eq:T}). Inset: (logarithm) of the effective coupling coefficient $\kappa_{\textup{eff}}$ (blue) and value of $\kappa_{\textup{eff}}^2=\kappa_{cc}^2$ at critical coupling (dashed violet line). (c) Extinction (blue) and $\lambda/\Delta\lambda$ (violet) of the target resonance $\lambda_0$. The  $\lambda/\Delta\lambda$ of an adjacent resonance at $\lambda_n \sim 1544.6$ nm is shown in green. (d) Stacked spectra of an adjacent resonance largely detuned from all Aux resonances.}
    \label{fig:picture2}
\end{figure*}

% The coupling coefficient in the Main resonator is set by the phase difference between the MZI arms, which is $\theta_{\textup{MZ}}=\theta_0+\theta_{\textup{aux}}$, where $\theta_{\textup{aux}}$ is the wavelength-dependent phase imparted to the light passing through the Aux resonator.
Initially, $\phi_{\textup{MZI}}$ is set to ensure that all the resonances of the Main resonator which are not perturbed by the Aux are critically coupled. This configuration mimics that shown in Fig.\ref{fig:IE}(d).
%Thereafter the spectra of the device while the auxiliary resonator is detuned were obtained. 
Figure \ref{fig:picture}(b) shows the device transmission when the Aux resonances are all detuned from those of the Main resonator. We can distinguish the Main from the Aux resonances from their different free spectral range (FSR),  linewidth and extinction. In particular, the resonances of low extinction are those of the Aux resonator because it is heavily overcoupled. %In Fig.\ref{fig:picture}(b) there are small peaks which are related to higher guided modes inside the resonators. 
The Main resonator has an FSR of $1.56$~nm. From the resonance linewidth in the critical coupling configuration we estimated an intrisic Q-factor of $\sim 3.2\times 10^6$ and $\kappa_A^2\sim 0.03$. The FSR of the auxiliary resonator is 2.1 nm and $\kappa_C^2 \sim 0.21$, yielding a Q of $\sim 23000$.
To demonstrate the change of the linewidth of a single resonance at $\lambda_0=1542.97$ (indicated with an arrow in Fig.\ref{fig:picture}(b)) in the Main resonator, we recorded the transmission spectra at different detunings $\Delta\lambda_{\textup{aux}}=\lambda_{\textup{Aux}}-\lambda_0$ between the target resonance and the closest auxiliary resonance at $\lambda_{\textup{Aux}}$. We then define the normalized distance between the two resonances as $\Delta\lambda_{\textup{aux}}/\gamma_{\textup{aux}}$, where $\gamma_{\textup{aux}} = 0.067$ nm is the linewidth of the Aux resonance. The wavelength of the Aux resonance is swept from $\Delta\lambda_{\textup{aux}}/\gamma_{\textup{aux}} = 5$ to $\Delta\lambda_{\textup{aux}}/\gamma_{\textup{aux}} \sim -2$ in small steps.  
The different spectra are shown stacked in Fig.\ref{fig:picture2}(a), while Fig.\ref{fig:picture2}(b) reports $T$ calculated using Eq.(\ref{eq:T}), which shows a good agreement with the experimental results. %Moreover, as an inset in Fig. \ref{fig:picture2}(b) the simulated effective coupling coefficient $\kappa^2_{\text{eff}}$ (in blue) and the critical coupling coefficient $\kappa^2_{cc}$ are shown as a function of the detuning. %In both of them we can distinguish the horizontal line of the resonance of the Main ring, meanwhile, the one that goes from left to right corresponds to the auxiliary. 
From the stacked spectra in Fig.\ref{fig:picture2}(a) we extracted the extinction and the linewidth $\Delta\lambda$ of the target resonance. In  Fig.\ref{fig:picture2}(c) it is shown the quantity $\lambda/\Delta\lambda$, which coincides with the Q-factor when the Aux is highly detuned from the target Main resonance, as the latter restores its natural Lorentzian shape. 
% The extinction and linewidth $\lambda/\Delta\lambda$ (which is equal to the Q factor when the Aux is highly detuned as the resonance has a Lorentzian shape) of the target resonance while the Aux is detuned are shown in Fig. \ref{fig:picture2}(c). 
The simultaneous analysis of the extinction and of the linewidth allows us to assess how the effective coupling $\kappa_{\textup{eff}}^2$ between the Main resonator and the bus-waveguide changes due to the phase imparted by the Aux resonator. When $\Delta\lambda_{\textup{aux}}/\gamma_{\textup{aux}}\sim 2.5$, the extinction is almost zero, and $\lambda/\Delta\lambda$ approaches the value of the intrinsic $Q$. In this configuration, the resonator is totally under-coupled at $\lambda_0$. This is also confirmed from the simulation of $\kappa_{\textup{eff}}^2$ as a function of  $\Delta\lambda_{\textup{aux}}/\gamma_{\textup{aux}}$ (inset in Fig.\ref{fig:picture2}(b)), where $\kappa_{\textup{eff}}$ reaches its minimum value at $\Delta\lambda_{\textup{aux}}/\gamma_{\textup{aux}}\sim 2.5$. At $\Delta\lambda_{\textup{aux}}/\gamma_{\textup{aux}}\sim 1.4$, the extinction is maximized ($\sim-10$ dB) and $\lambda/\Delta\lambda$ is half the intrinsic Q. This implies that the resonance is critically coupled, meaning that $\kappa_{\text{eff}}=\kappa_{cc}$ at $\lambda_0$, as shown in the inset of Fig.\ref{fig:picture2}(b). When $\Delta\lambda_{\textup{aux}}/\gamma_{\textup{aux}}\rightarrow 0$, we have that $\lambda/\Delta\lambda$ reaches the minimum value of $6.5\times10^4$, with an extinction of $-0.25$ dB, indicating an highly over-coupled resonance which is consistent with the maximum in $\kappa^2_{\text{eff}}$ shown in the inset of Fig.\ref{fig:picture2}(b).

To summarize, by sweeping the Aux resonance across the target resonance of the Main, we can continuously control $\lambda/\Delta\lambda$ between $6.5\times10^4$ and $\sim 3\times 10^6$, realizing all the coupling regimes described in Sec.\ref{sec:theory}.
To confirm that the control of the Q-factor is wavelength-selective, in Fig.\ref{fig:picture2}(c) we also report the linewidth of a resonance at $\lambda_n\sim 1544.6$ nm, which is largely detuned from all Aux resonances. As  $\Delta\lambda_{\textup{aux}}$ changes, the Q-factor reMains mostly unperturbed. 
The stacked spectra around the resonance at $\lambda_n$ are shown in Fig.\ref{fig:picture2}(d). The slight red-shift of the resonance wavelength is caused by the thermal cross-talk between the micro-heater on top of the Aux resonator and the Main resonator.

\section{\label{sec:concl}Conclusions} 
We proposed and experimentally validated an integrated photonic device consisting of a microresonator where the Q-factor of a single resonance can be selectively changed, leaving that of the other resonances unperturbed. This is accomplished by a resonant interferometric coupler, in which an auxiliary resonator allows us to change the interference condition in a very narrow-band spectral range. We theoretically investigated several configurations, in which one  resonance of the Main resonator is tuned from being maximally over-coupled to be totally under-coupled, leaving the other resonances unchanged.   
We experimentally validated the working principle of the device on the silicon nitride platform. The fabricated device enables the selective switching of the quality factor of a target resonance from $6.5\times 10^4$ to $3\times 10^6$. 
By adding more than one Aux resonator in the long arm of the MZI we could target different resonances and independently control them. Thus, our design is highly modular and scalable.
Many applications that require the resonant control of the light matter interaction in nonlinear and quantum optics could benefit from our structure. Examples includes Q-switched lasers, wavelength-selective integrated parametric oscillators, entangled photon pair generation and photon pair sources with simultaneously high brightness heralding efficiency.
%Paula's version
%Other applications, could be use it to reach the maximum photon-pair generation rate, combination of a critically coupled resonance for the pump and over-coupled resonances for the signal and idler \cite{chen2023pushing}. 
%Alice's version
%Another application could be to maximize the photon-pair generation rate by combining a critically coupled resonance for the pump and over-coupled resonances for the signal and idler \cite{chen2023pushing}.  
\section*{Acknowledgements}
P.L.P. acknowledges funding from the HyperSpace project (project ID 101070168). D.B acknowledges the support of Italian MUR and the European Union - Next Generation EU through the PRIN project number F53D23000550006 - SIGNED. M.B., M.G and M.L. acknowledge the PNRR MUR project PE0000023-NQSTI. All the authors acknowledge the support of Xanadu Quantum Technology for providing the samples and the useful discussions.

\vspace{1cm}

%\begin{acknowledgments}
%We wish to acknowledge the support of the author community in using
%REV\TeX{}, offering suggestions and encouragement, testing new versions,
%\end{acknowledgments}

% \appendix

% \section{Appendixes}

%\nocite{*}

%\bibliography{Biblio}% Produces the bibliography via BibTeX.

\end{document}